# Size dependence of the Graphene Islands Moving on Cu (111) Surface during the CVD Growth

*Ziwei Xu,\*[†] Changshuai Shi,[†] Lu Qiu,[‡] Feng Ding\*[‡]*

[†] School of Materials Science & Engineering, Jiangsu University, Zhenjiang 212013, China

[‡] Centre for Multidimensional Carbon Materials, Institute for Basic Science, Ulsan 44919, Korea

**ABSTRACT**

The graphene islands, formed as different sizes, are crucial for the final quality of the formed graphene during the CVD growth either as the nucleation seeds or as the build blocks for larger graphene domains. Extensive efforts had been devoted to the size or the morphology control while fewer works were reported on the moving dynamics of these graphene islands as well as the associate influences to their coalescence during the CVD Growth of graphene. In this study, based on the self-developed C-Cu empirical potential, we performed systematic molecular dynamics simulations on the surface moving of three typical graphene islands $C_N$ (N = 24, 54 and 96) on the Cu (111) surface and discovered their different behaviors in sinking, lateral translation and rotation at the atomic scale owing to their different sizes, which were proved to bring forth significant impacts to their coalescences and the final quality of the as-formed larger domains of graphene. This study would deepen our atomistic insights into the mechanisms of the graphene CVD growth and provide significant theoretical guidelines to its controlled synthesis.

# Introduction

Since the first exfoliation in 2004, graphene has gathered great attentions both in fundamental and technological explorations owing to its unique properties in morphologies, mechanics, electronics, optics, and so on. For many applications, wafer-scale single-crystalline graphene films are extremely expected as an ideal platform for the design of future high-performance graphene based devices.[1] Great efforts, consequently, have been paid to synthesize wafer-scale single-crystalline graphene monolayers[2-4] on the transition metals through chemical vapor deposition (CVD) growth--the most promising, cheap and readily accessible approach.[5] Among all of the metals, the copper[6-8] are proved as an ideal substrate for the synthesis of large-area high-quality monolayer graphene.[2-3, 9] Different from the nickel substrate, the growth of graphene on the copper is a surface mediated process due to the low solubility of the carbon atoms in the bulk of copper substrate. Most recently, fast growth of wafer-scale single-crystalline graphene monolayers was realized either by controlling the nucleus[8] or by the continuous oxygen supply on the Cu (or Cu-Ni alloys)[7].

It is well known that the graphene islands formed during the initial nucleation[10-15] are crucial for the final quality of the grown graphene either as the nucleation seeds [10, 16-17] or as the build blocks for coalescence the graphene nano-islands.[13] Experiments discovered that the diffused graphene islands can coalescence seamlessly on the liquid copper surface[6] or the hydrogenated hydrogen-terminated germanium surface.[18] During the initial stage, however, various carbon intermediates, like carbon dimers, atomic-carbon nanoarches carbon clusters with networks, were discovered both in the experiments [19-20] and theoretic calculations.[21-23] The unexpected defects, such as pentagon, were usually introduced during the nucleation stage, which degrades the final quality of the grown graphene. To circumvent these detrimental defects created at the initial stage, the defect-free polycyclic aromatic hydrocarbons (PAHs), like containing the six membered carbon rings, were put forward as the ideal precursors for the low-defect graphene fabrication on Au (111)[24], Pt (111)[25] and Cu (111).[26] Among all of those motifs, the coronene-like cluster, $C_{24}$, was widely regarded as the ideal carbon precursor for the self-assembly growth of high-quality graphene on Cu (111) surface,[27-28] which has 6-foled rotation symmetry with dome-shaped structure induced by the interaction between the peripheral atoms and the substrate metals.[29-31]

Although it is widely believed that these active carbon species moved on the metal surface, coalesced into bigger islands and eventually formed the final graphene,[27, 32] the moving and calescence dynamics of these carbon islands were still elusive at the atomic level. For example, a new growth mode, named as "embedded" growth mode has been proposed based on the static DFT calculations recently. In this embedded mode, it is energetically favored that the graphene islands tend to "sink" into the "soft" metals like Au (111), Pd (111) and Cu (111) during the growth of graphene.[33] The sinking process, however, has never been proved based on the molecular dynamics (MD) simulation. Hence, a deep investigation to the moving dynamics of graphene islands on the metal surface is hence highly desired for the better controlled synthesis of large-area and high-quality graphene. In this article, based on the self-developed C-Cu potential, a classical MD simulation were performed to study the moving dynamics of three grapheme islands with different sizes ($C_{24}$, $C_{54}$ and $C_{96}$) on the Cu (111) surface, including their sinking, lateral translations as well as rotations. A clear dependence of these moving dynamics on their sizes was discovered with the associated mechanisms analyzed, which were expected to bring forth significant impacts to their coalescences and the final quality of the as-formed graphene domains.

## Results and Discussion

## Sinking

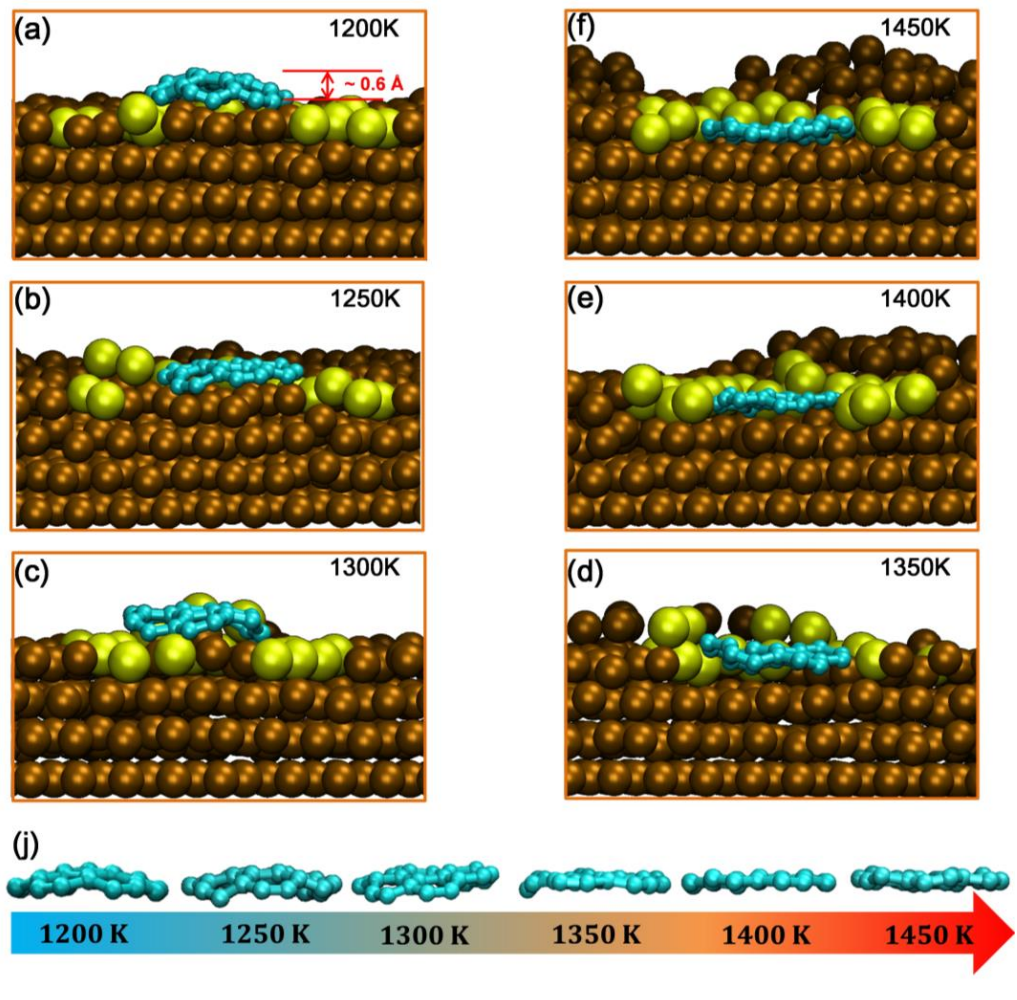

**Figure 1.** (color online). Temperature effects on the sinking of $C_{24}$ on the Cu (111) surface. (a-f) Cross-sections of the $C_{24}$ on the Cu (111) surface after 120 ps MD simulations at 1200 K, 1250 K, 1300 K, 1350 K, 1400 K and 1450 K, respectively. (j) The profiles of $C_{24}$ extracted from (a-f) at different temperatures. The cyan and ochre balls represent the Carbon and Cu atoms, respectively. The Cu atoms near the edge of the $C_{24}$ are highlighted by yellow.

Figure 1 shows the cross-sections of the $C_{24}$ on the Cu (111) surface after 120 ps MD simulations for different temperatures (1200 K → 1450 K). The full views of the corresponding configurations are shown in Fig. S2. As shown in Figure 1a-b (Fig. S2a-b), the Cu (111) surface is stable in retaining its original crystallinity at the lower temperature (T < 1300 K). The $C_{24}$ is floating on the surface during the whole simulations with invisible sinking spotted. Previous studies discovered that the graphene edge favored attaching to the metal step during the graphene growth with the great reduction of the edge formation energy.[10, 12, 34] Due to the dominance of metal-carbon σ bonding over metal-carbon π bonding, the graphene ribbon tended to stand uprightly on the flat metal terrace[35] while the graphene precursor like $C_{24}$, is likely to form domelike structure with the edge bended toward to the terrace.[21] As shown in Fig. 1a, the $C_{24}$ also presents a domelike geometry like the previous studies[21] with the edge bended to the Cu (111) surface. Owning to the weaker bonding between the edge of $C_{24}$ and the Cu (111) surface, the height difference between the top and bottom of this cap is ~ 0.6 Å, which is ~ 50% of that calculated from the optimized geometry on the Ni (111) substrate.[36] At 1300 K, the Cu (111) surface starts to melt slightly with minor atoms of Cu diffusing out of the surface and attaching to the edge of $C_{24}$ (Fig. 1c, S1c). These diffused atoms of Cu can be viewed as a new-formed atomic layer passivating the edge of $C_{24}$. Therefore, unlike the domelike flake on the Cu surface shown in Fig.1a, the $C_{24}$ can form metal-carbon σ bonding with flatter edge instead of bending toward to the surface (Fig. 1c, j). At the temperature higher than 1300 K, the degree of the surface melting increases with more disordered area appearing. Obviously, such a surface melting makes the sinking of $C_{24}$ easier to take place. As shown in Fig. 1d-f, the $C_{24}$ sinks down and embeds itself in the melted Cu surface at the temperature of 1350 K, 1400 K and 1450 K, respectively. The higher the temperature is, the deeper the $C_{24}$ sinks down. Besides, with the further sinking, the domelike $C_{24}$ became flatter and flatter (Fig. 1j) at the higher temperatures.

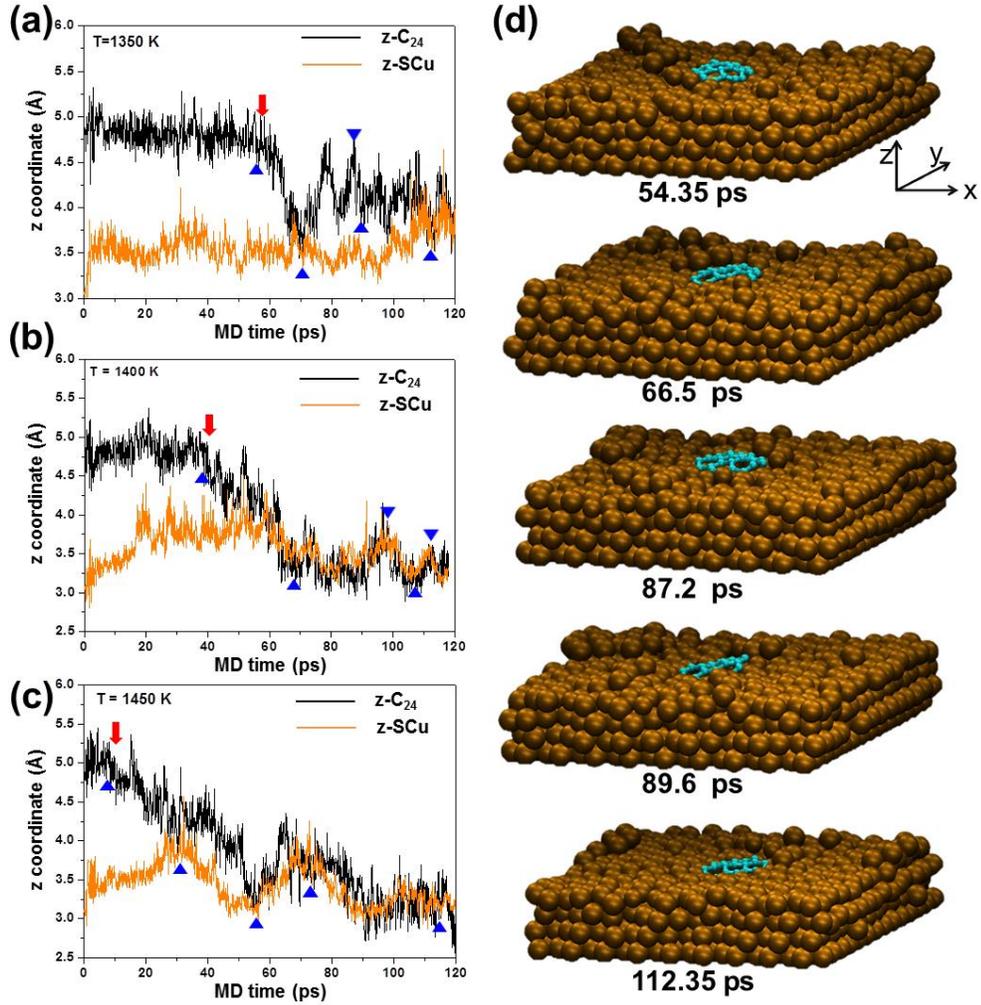

**Figure 2.** (color online). (a-c) Heights of the $C_{24}$ cluster (z-$C_{24}$) and the surrounding Cu atoms (z-SCu) during the MD simulation of the $C_{24}$ on Cu (111) surface at 1350 K, 1400 K and 1450 K, respectively. The black line and yellow line denote the z-$C_{24}$ and z-SCu, respectively. The red arrows represent the starting time of the sinking. The blue arrows represent the snapshots extracted during the MD. (d) Snapshots extracted from the MD trajectories of the $C_{24}$ on Cu (111) surface at 1350 K. The corresponding z-$C_{24}$ and z-SCu of these extracted snapshots are denoted as the blue arrows in (a). The cyan and ochre balls represent the Carbon and Cu atoms, respectively.

Hence, the driving force for the sinking is the energetic compensation of the curvature energy release (domelike flake → flat flake) as well as a better edge passivation by the metal step formed in the sinking basin (Fig. 1d-f). Yuan et. al., proposed that the sunk-mode of the graphene

CVD growth can be achieved by two possible ways: i) by removing the metal atoms around graphene island and ii) by adsorbing the fast migrating metal atoms around the graphene island to form a new atomic layer.[33] For the $C_{24}$, clearly, the former one is the dominant sunk-mode at the temperatures higher than 1300 K.

In order to quantify the degree of the $C_{24}$ sinking, we calculate the z coordinates of the mass centers of $C_{24}$ (z-$C_{24}$) and its surrounding Cu atoms (z-SCu) (see the axes in Fig. 2d) to reflect the variations of their heights as a function of MD time. As shown in Fig. S1, at the temperatures lower than 1300 K, the two lines are almost parallel implying that neither the sinking of $C_{24}$ nor the surface melting is able to occur (Fig. S1a-b). At 1300 K, both the z-$C_{24}$ and z-SCu start to rise after 90 ps. The reason for such a rise can be mainly attributed to the change of the mass centers of the $C_{24}$ and the surrounding Cu atoms. In another word, the original domelike $C_{24}$ at the lower temperature begin to flatten at 1300 K due to the edge passivation by several Cu atoms diffused up. Therefore, the mass center of the $C_{24}$ is raised by ~0.3-0.5 Å, which is consistent with the previous density function theory (DFT) calculation on the height difference between $C_{24}$ and the coronene on the metal substrate.[21] In addition, the mass center of the surrounding Cu atoms is also raised by ~0.5-0.75 Å because of the diffusing up of the Cu atoms. Up to now, the $C_{24}$ is floating on the Cu (111) surface without sinking. As shown in Fig. 2(a), however, when the temperature is at 1350 K, the z-$C_{24}$ begins to drop sharply after ~66.5 ps (red arrow) and overlap the z-SCu during the ~65ps - 75ps and ~97ps -120 ps, indicating the $C_{24}$ has fully sunk into the melted Cu surface (Fig. 2d). With the further increase of the temperature, the onset time of the $C_{24}$ sinking, denoted by the red arrows, is advanced from 60 ps to 40 ps (T=1400 K) and then to 10 ps (T=1450 K), respectively (Fig. 2b,c). Moreover, unlike the intermittent overlapping at 1350 K, the lines of z-$C_{24}$ and z-SCu are fully overlapping almost during the whole MD time after the sinking occurs. Nevertheless, with the increase of the temperature, both two lines wave synchronously from 60-100 ps, which reflect the rolling surface of molten Cu induced by such a high temperature while the $C_{24}$ is just like a sinking "boat" on the rolling "sea". Fig. 3a shows the mean values of z coordinates, $\bar{z}$-$C_{24}$ and $\bar{z}$-SCu, during the final stage (~ 5 ps) for different temperatures. The $\bar{z}$-$C_{24}$ displays a large decrease with the increase of temperature whereas the variation of $\bar{z}$-SCu is negligible with only a slight fluctuation of around 3.5 Å. Fig. 3b shows that their difference $\Delta\bar{z}$, namely $\Delta\bar{z} = \bar{z}$-$C_{24}$—$\bar{z}$-SCu, drops below 0 Å at the temperatures higher than 1350 K indicating the complete sinking of the $C_{24}$.

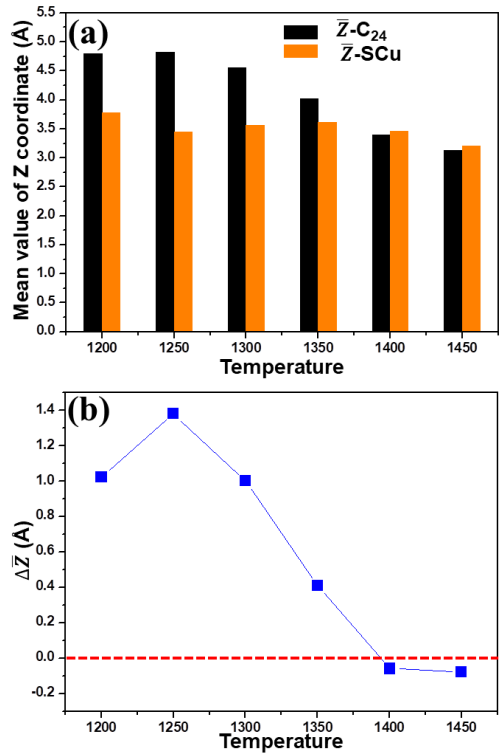

**Figure 3.** (color online). (a) $\bar{z}$-$C_{24}$, $\bar{z}$-SCu and (b) $\Delta\bar{z}$ vs. temperature. The dashed red line denotes $\Delta\bar{z}=0$.

After knowing the sunk-mode of the $C_{24}$ at the higher temperatures, we may ask one question consequently: Does the larger graphene island also present the similar sunk-mode? To answer this question, the carbon clusters of $C_{54}$ and $C_{96}$, including 19 and 37 hexagons, on the center of Cu (111) surfaces were selected for the further simulations. As shown in Fig. 4a, unlike the $C_{24}$, no obvious sinking down of the $C_{54}$ is seen at 1350 K. The z-$C_{54}$ and z-SCu in Fig. 4d also prove that there is a big gap of the distance along the z direction between $C_{54}$ and the Cu surface below. Even at the higher temperatures like 1400 K and 1450 K (Figs. 4b and c), the $C_{54}$ prefers to float on the Cu surface and the surface Cu atoms eventually melt and diffuse up to surround the $C_{54}$. Although the gap between the z-$C_{54}$ and z-SCu is reduced in Fig. 4(e-f), the reason can only be attributed to the up-lift of the surrounding Cu atoms instead of the sinking down of the $C_{54}$. In another word, the $C_{54}$, cannot sink down on the Cu (111) surface as easily as the $C_{24}$ does. The reason is that the larger size of the $C_{54}$ costs more energy by kicking out the Cu atoms below. Besides, the $C_{54}$ is more energetic stable on the Cu (111) surface owning to the smaller ratio of peripheral edge atoms/inner atoms. Thus, the driving force for the sinking down of the $C_{54}$ is hence smaller than the one of $C_{24}$. In this respect, the sunk-mode of the $C_{54}$ at the high temperatures can be categorized into the later one as mentioned above. As for the larger one, like $C_{96}$, it is always floating on the Cu (111) surface from 1300 K to 1450 K (Fig. S4). We can hence conclude that the larger the carbon nanocluster, the harder the sinking occurs.

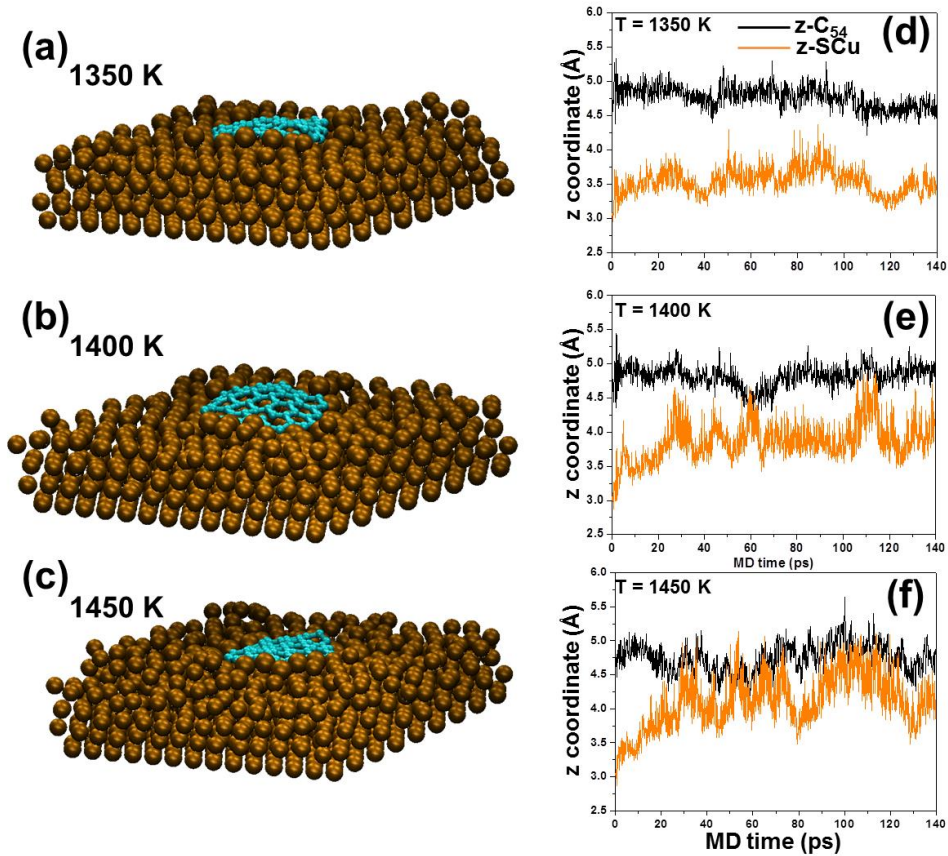

**Figure 4.** (color online). (a-c) Final structures of the $C_{54}$ on Cu (111) surface during the MD at 1350 K, 1400 K and 1450 K, respectively. The cyan and ochre balls represent the Carbon and Cu atoms, respectively. (d-f) z-$C_{54}$ and z-SCu vs. MD time at 1350 K, 1400 K and 1450 K, respectively.

## Translation & Rotation

In addition to the sinking behavior, the translation and the rotation of the graphene islands at the surface of Cu are also significant factors reflecting the migration dynamics during their coalescence. Figures 5 and S4 show the coordinate (x, y) trajectories of the $C_{24}$, $C_{54}$ and $C_{96}$, respectively, on Cu (111) surfaces at different temperatures during the MD. Very strikingly, the trajectory patterns of the translation are largely influenced by the cluster sizes as well as the temperatures, reflecting their different surface-diffusion abilities. As shown in Figs. 5 and S5, the translation trajectory of $C_{24}$ at the temperature ≤ 1200 K displays a single dispersive "cloud" indicating that the isotropic rigid vibration of the $C_{24}$ at the surface due to its isotropic edges (armchair edges). The maximal translation ranges along x ($MTR_x$) and y directions ($MTR_y$) are only among 1 and 1.5 Å, respectively, which rise slightly with the increase of the temperature. When the temperature rises to 1250 K, the $MTR_x$ and $MTR_y$ of $C_{24}$ start to increase dramatically with the single dispersive "cloud" replaced by multiple connected dispersive "clouds". The higher the temperature, the more dispersive "clouds" are. These multiple "clouds" of trajectory are eventually connected to be larger one at higher temperatures of 1300 K, 1350 K, 1400 K and 1450 K, respectively, indicating the long-range diffusion of the $C_{24}$. On the other hand, the $C_{54}$ displays quite different translation trajectories. As shown in Figs. 5 and S5, at the low temperature like 600 K or 800 K, the migration trajectory presents directional drift instead of the single dispersive "cloud", which leads to larger $MTR_x$ and/or $MTR_y$ than those of $C_{24}$ at the same temperature. The reason for such different migration trajectories may be attributed to the alternative armchair (AC) and zigzag (ZZ) edge atoms along the circumference of $C_{54}$. At the low temperature, the difference of the carbon-metal interactions between AC and ZZ sites make the isotropic diffusion is impossible. Assisted by the vibration of Cu and carbon atoms, the mass center of the $C_{54}$ undergoes a collective drift along an initial direction. For such a reason, the $MTR_x$ and/or $MTR_y$ are larger than those of the $C_{24}$. When the temperature is at 1000 K, however, more kinetic energies are provided to smooth the anisotropic diffusion at the lower temperature. For such a reason, the directional drift of $C_{54}$ disappears once the temperature increases. At the higher temperature $\geq$ 1350 K, the trajectory of the $C_{54}$ is also connected by multiple "clouds" indicating the long-range translation occurs, which is similar to that of $C_{24}$. As for the $C_{96}$, very strikingly, the trajectory patterns were largely degenerated to the single dispersive "clouds", even at the higher temperatures. Such specific patterns for $C_{96}$ can be ascribed to the higher diffusion damping's created by the larger isotropic peripheral edge dominated by the zigzag sites, which suppressed the lateral diffusion of $C_{96}$ as only a small vibration with respect to its mass center.

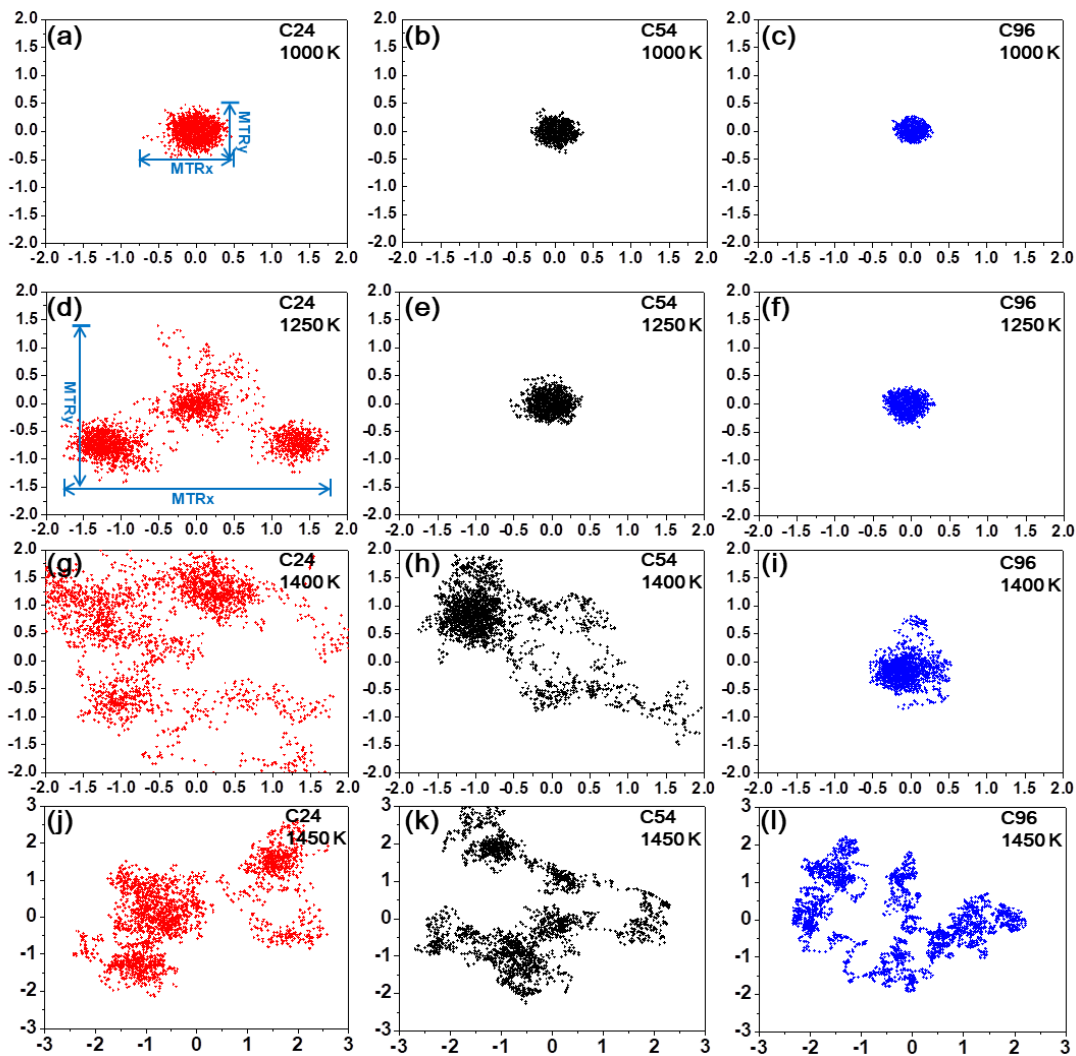

**Figure 5.** (color online). The coordinate (x, y) trajectories of the lateral translations for $C_{24}$, $C_{54}$ and $C_{96}$ on Cu (111) surfaces, respectively, at different temperatures. The maximal translation ranges along x and y directions (MTRx and MTRy) are defined in (a) and (d).

In order to compare the migration abilities of these three grphene islands, we presented the $MTR_x$ and $MTR_y$ as a function of temperature in Fig. 6 for $C_{24}$, $C_{54}$, and $C_{96}$, respectively. When the T < 1000 K, the $MTR_x$ and $MTR_y$ of $C_{54}$ are larger than those of $C_{24}$, particularly at T = 800 K. Such a larger $MTR_x$ and $MTR_y$ is simply caused by the directional drift of $C_{54}$. In contrast, it should be noted that the MTR$_x$ and MTR$_y$ of the C$_{54}$ are smaller than those of C$_{24}$ within 1000 K ~ 1400 K. Such a result is caused by the larger damping induced by the larger peripheral edge atoms belonged to C$_{54}$. At the temperature 1450 K, the MTR$_x$ and MTR$_y$ of C$_{24}$ drop distinctly, which was induced by the complete sinking of the C$_{24}$ into the melted Cu (111) surface. The diffusion ability is hence largely weakened. Compared with C$_{24}$ and C$_{54}$, the MTRs of C$_{96}$ is smallest at all temperatures. Moreover, the C$_{96}$ is rigidly pinned until 1400 K, indicating that if the diameter of the carbon nanocluster is larger than 20 Å, its diffusion on the Cu (111) surface would be greatly suppressed.

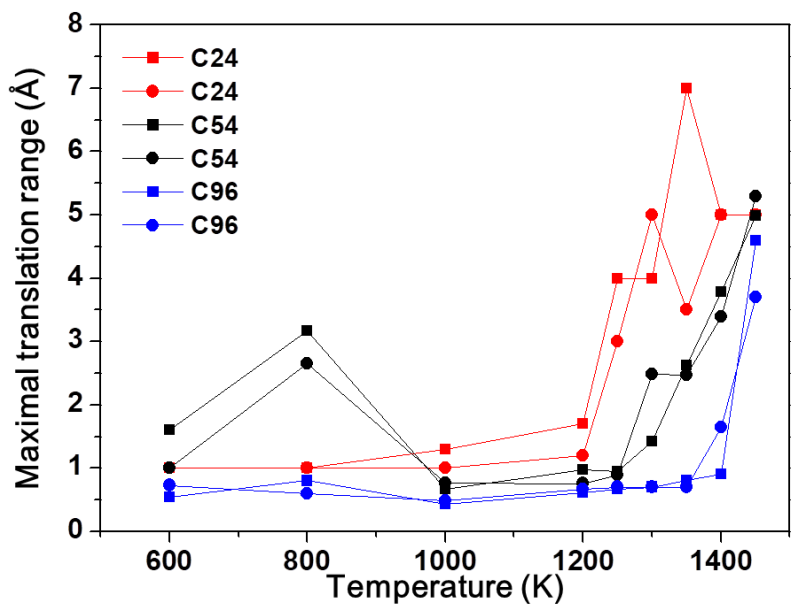

**Figure 6.** (color online). Maximal translation ranges (MTRs) along x and y directions (MTRx and MTRy) for C$_{24}$, C$_{54}$ and C$_{96}$ at different temperatures. The MTRx and MTRy are represented by the solid rectangular and circle, respectively.

Besides vertical sinking and lateral translation, the lateral rotation with respect to the mass center of the carbon flake is also a crucial factor determining the later growth of the graphene, such as the orientation, grain boundary and so on. Figures 7 and S5 show the rotation angles θ vs. the MD times for C$_{24}$, C$_{54}$ and C$_{96}$ at different temperatures, respectively. At the temperature ≤ 1300 K, the $C_{24}$ displays a rigid vibration at θ = 0°. Occasionally, it rotates with the θ ~ 20° but then returns to θ = 0 ° very soon. These occasional rotations result in many sharp peaks consequently at temperature ≤ 1300 K. The higher the temperature, the more peaks appear. At 1350 K, however, the $C_{24}$ starts to rotate from 20 ps to 70 ps with the θ ~ 80 ° reached. After that, the θ returns back to 60 °. Interestingly, when the temperature increases to 1400 K or 1450 K, the onset time for the large irreversible rotation is somehow delayed to 100 ps or 90 ps (red arrows), which is much later than that at 1350 K. The magnitude of the irreversible rotation angle θ is ~ 40 °, also smaller than the one at 1350 K. This delay might be attributed to the increased sinking degree of the $C_{24}$ aforementioned. In another word, at 1350 K, the $C_{24}$ has not sink into the Cu atoms completely compared with that at 1400 K or 1450 K. The damping of the rotation is hence smaller. For the larger carbon flake of $C_{54}$, on the other hand, its rotation behaviors are quite different. As shown in Fig. 7, for instance, at the temperature ≤ 1300 K, the sharp peaks are much weaker than those of $C_{24}$. Even at the temperature of 1350 K, the rotated angle is also smaller than that of the $C_{24}$. Such a reduction of the rotation ability of $C_{54}$ can also be ascribed to the larger damping created by the larger and anisotropic peripheral edge of $C_{54}$, which is mixed with armchair and zigzag carbon atoms. Nevertheless, since the $C_{54}$ prefers to float on the melted Cu surface rather than sink down at the higher temperatures like 1400 K and 1450 K, the rotation damping encountered is conversely smaller compared with that encountered by $C_{24}$. Thus, the magnitude of the rotation θ increases when the temperature rises to 1400 K and 1450 K. The onset times of irreversible rotation of $C_{54}$ at 1350 K, 1400 K and 1450 K are also advanced to 40 ps, 30 ps and then to 5 ps (red arrows), respectively, which is quite different from that of $C_{24}$. As for the larger cluster of $C_{96}$, its rotation is also greatly suppressed. As shown in Fig. 7, the $C_{96}$ almost remained stationary without remarkable rotation until 1450 K.

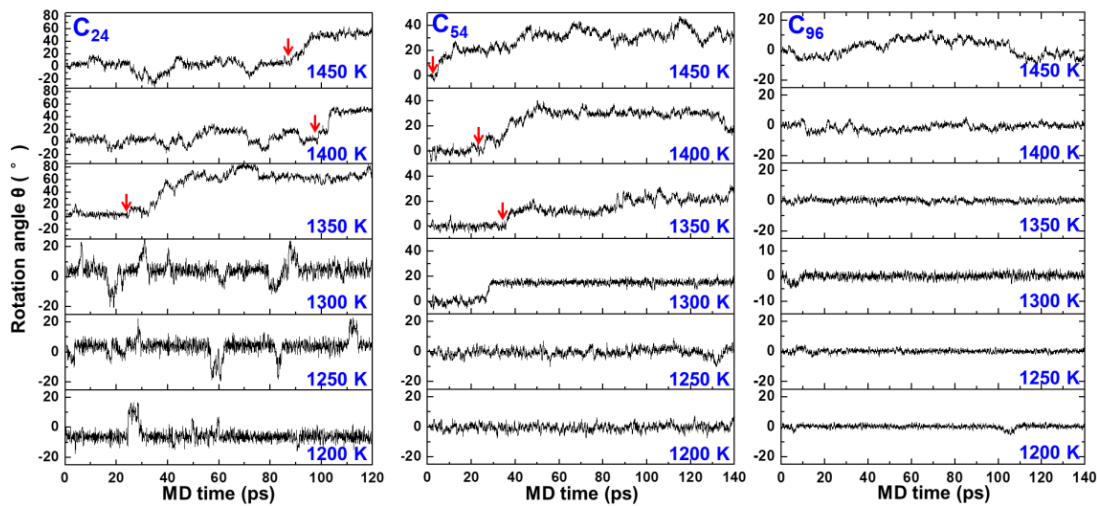

**Figure 7.** (color online). Rotation angles θ vs. the MD times for $C_{24}$ (left), $C_{54}$ (middle) and $C_{96}$ (right) at different temperatures. The onset times of the large irreversible rotations are detonated by red arrows for $C_{24}$ and $C_{54}$, respectively.

## Coalescence of the graphene islands

After making clear the moving dynamics of the isolated graphene islands, let's turn to their coalescence that leads to the formation of the larger graphene domain. Figure 8 displays the final structures of eight $C_{24}$s on the Cu (111) surface at different temperatures after 2 ns MD simulations. As expected, the coalescence degree of these $C_{24}$s depends largely on the temperature. At the very low temperature of 600 K, for instance, no visible translation and rotation is seen except for the minor vibrations (Fig. 8a). The increased temperature is able to promote the translation and the rotation of the $C_{24}$s on the Cu (111) surface and their coalescence is consequently accelerated (Fig. 8b-f). As shown in Fig. 8b, those $C_{24}$s started to coalesce had changed their initial orientations with small rogations at lower temperature of 800 K (Fig. 8b), which was driven by their chemical bonding. In contrast, those isolated $C_{24}$s still remained their initial orientations. Furthermore, it is worth noting that the sinking down of the $C_{24}$ during the coalescence occurs at the temperatures of 1350 K and 1400 K, respectively. In fact, during the coalescence, one of the $C_{24}$ has already presented the tendency of sinking down at the temperature of 1200 K (Fig. 8d). At the temperature of 1350 K (or 1400 K), the sinking down of the $C_{24}$ appears with one layer of Cu atoms removed completely (Fig. 8e, f). This result agrees very well with the sinking dynamics of isolated $C_{24}$ uncovered above. On the contrary, until 1350 K, all of the $C_{54}$s (Fig. S7) are floating on the Cu (111) surface during the coalescence without sinking down. Instead, the melted Cu atoms diffuse out and form a new adlayer passivating the edges of the coalesced $C_{54}$, which is the so-called second sunk-mode. For this reason, the graphene islands coalesced by $C_{54}$s at the higher temperatures are flatter than those of $C_{24}$s.

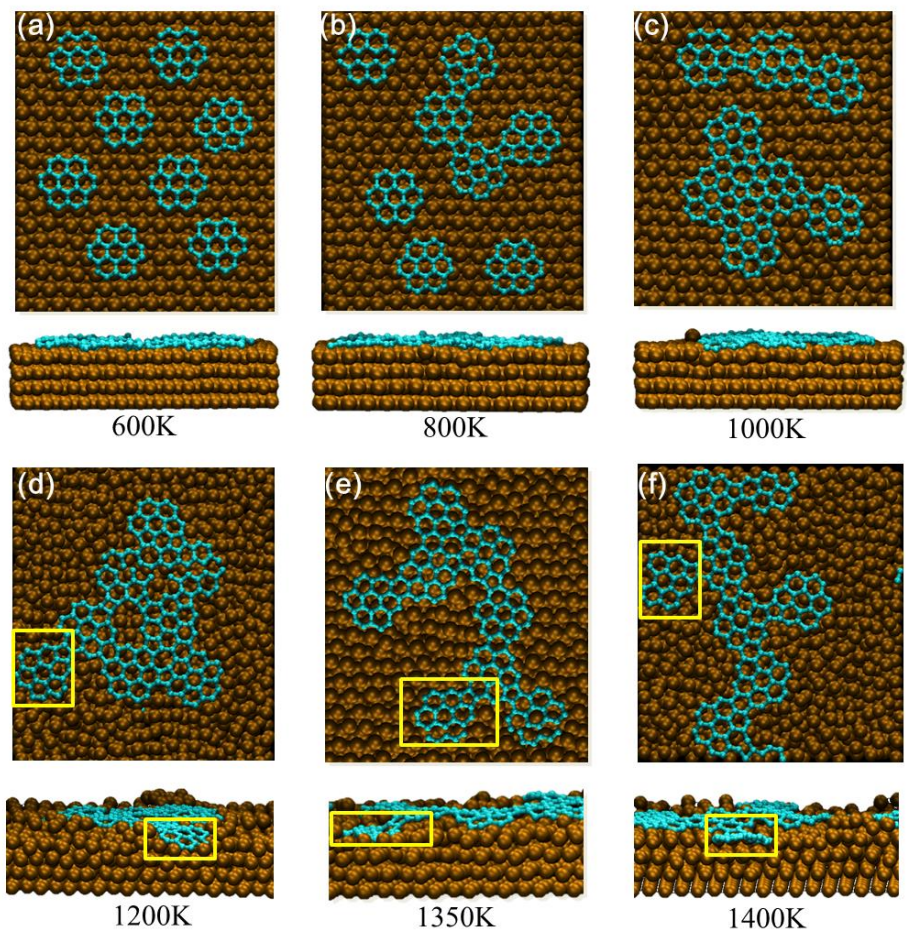

**Figure 8** Final structures of the coalesced $C_{24}$s on Cu (111) surface after 2 ns MD simulations at different temperatures. The cyan and ochre balls represent the Carbon and Cu atoms, respectively. The sinking down of the $C_{24}$ during the coalescence is highlighted by the yellow rectangle.

**CONCLUSIONS**

In conclusion, based on the classical MD simulation, the moving dynamics of the graphene islands $C_N$ (N=20, 54 and 96) on the Cu (111) surface were systematically investigated at the atomic scale. In details, we compared the sinking, lateral translation and lateral rotation of these clusters and discovered that: 1) The sinking of the carbon clusters occurred at the higher temperature (T > 1200 K ~ 1300 K) but with different modes (or say ways) depending on the size as well as edge of the cluster. For instance, the $C_{24}$ could really sinking down to the sub-layer of the Cu surface below while the $C_{54}$ or the larger one preferred to float on the surface and form the so-called "sunk" with its edge atoms passivated by the melted Cu atoms, which actually formed a new atomic layer on the original Cu surface . 2) The larger graphene islands like $C_{54}$ and $C_{96}$ had lower lateral translation & rotation ability than those of $C_{24}$ at the higher temperature (T ≥ 1000 K) due to the larger damping created by their edges. Besides, the sinking modes also affected the translation & rotation abilities at the higher temperatures. 3) During the coalescence of the multiple carbon clusters, the moving dynamics of the isolated ones, such as sinking, translation and rotation, could determine the coalescence degree and the quality of the as-formed graphene domains. Although the higher temperature is able to facilitate the coalescence degree of these merged carbon clusters, the sinking down of the $C_{24}$ may deteriorate the quality of the as-formed graphene domain. In contrast, the larger $C_{54}$ can avoid the sinking down at the higher temperature and form flatter and better graphene domain.

## COMPUTATIONAL METHODS

A four-layer rectangular Cu (111) slab with the periodicity of 35.75 Å x 39.84 Å x 100 Å was adopted as the catalyst substrate. To mimic the semi-infinite surface, the Cu atoms in the bottom layer were fixed during the simulation. The grphene islands of $C_{24}$, $C_{54}$ and $C_{96}$ were placed on the center of the Cu (111) surface and were optimized as the initial structures of the MD simulations. The pure classical molecular dynamics (MD) simulations were performed with the time step of 0.5 fs. Just like our previous work,[37] the velocity Verlet algorithm and the Berendsen thermostat[38] was adopted in the MD. The fundamental formats of the empirical potential energy surface (PES) employed in this classical MD were mainly based on the new generation potential of carbon-metal interaction that had already been successfully applied in our previous simulations on the growths of the carbon nanotube and graphene.[36-37] Specifically, the fundamental format of this new PES was revised mainly based on the modified second generation of the reactive empirical bond order ($REBO^2$) potential for the carbon-carbon interactions,[39] the pioneer potentials for the carbon-metal interactions[40-42] and the many-body tight-binding (TB) potential for metal–metal interactions[43] with 26 new added parameters for the further tuning of the carbon-metal interactions under various chemical environments, in particular the angle-dependent edge-substrate interaction of graphene.[36] In this article, new parameters fittings were performed to describe the C-Cu potential. The benchmark structures for the parameters fittings of C-Cu system were similar to those selected in the previous work [36] with only the metals Ni replaced by Cu. The energies for the parameters fittings listed in Table S1 and S2, which were set according to the formation energies of those benchmark structures calculated by the density functional theory (DFT). Particularly, the grphene islands of $C_{20}$, $C_{21}$ and $C_{24}$ were also adopted as the test structures for the new fitted PES of C-Cu system. Their formation energies calculated by the new PES and the DFT were listed in Table S3, which were very close.

*Conflict of Interest*: The author declares no competing financial interest.

*Acknowledgement:* This work is supported by the National Natural Science Foundation of China (11774136, 11404144, 51572111), Project funded by China Postdoctoral Science Foundation (2016M601722, 2018T110445), the Advanced Talents Foundation of Jiangsu University (14JDG120). The authors are thankful for the High Performance Computing Platform of Jiangsu University.

*Supporting Information Available:* See supplementary material for Table S1-S3; Fig. S1- S6.

*\* Correspondence author:*ziweixu2014@ujs.edu.cn; feng.ding@polyu.edu.hk

**Graphic Table of Contents**

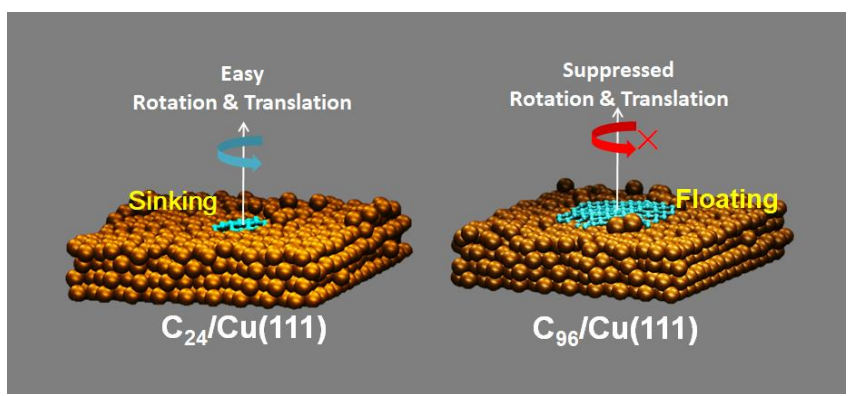

The migration dynamics of the graphene cluster on the Cu(111) surface, including sinking, rotation and translation, are significantly influenced by the cluster's size.